\documentclass[%
 notitlepage,
 twocolumn,
superscriptaddress,
nofootinbib,
 amsmath,amssymb,
 aps,
 pra,
]{revtex4-2}
\usepackage{amsfonts}
\usepackage{graphicx}
\usepackage{physics}
\usepackage{pgf} 
\usepackage{dcolumn}
\usepackage{bm}
\usepackage{bbold}
\usepackage{soul}
\usepackage{cancel}
\usepackage{latexsym}
\usepackage{color}
\usepackage{xcolor}
\usepackage{dsfont}
\usepackage{enumerate}
\usepackage{comment}
\usepackage{ulem}
\usepackage{hyperref}
\hypersetup{
citecolor=blue,
colorlinks=true,
urlcolor=magenta,
}

\newcommand{\IOPPAS}{Institute of Physics PAS, Aleja Lotnikow 32/46, 02-668 Warszawa, Poland}

\newcommand{\ENS}{Laboratoire Kastler Brossel, ENS-Universit\'e PSL, CNRS, Sorbonne Universit\'e, Coll\`ege de France, 24 rue Lhomond, 75005 Paris, France}

\newcommand{\UIBK}{Universität Innsbruck, Fakultät für Mathematik, Informatik und Physik, Institut für Experimentalphysik, 6020 Innsbruck, Austria}

\begin{document}

\title{
Bounds on detection of Bell correlations with entangled ultra-cold atoms in optical lattices under occupation defects
}

\author{T. Hernández Yanes$^*$}
\affiliation{\IOPPAS}
\affiliation{\UIBK}

\author{Y. Bamaara$^*$}
\affiliation{\ENS}

\author{A. Sinatra}
\affiliation{\ENS}

\author{E. Witkowska}
\affiliation{\IOPPAS}

\date{\today}

\begin{abstract}
Bell non-locality stems from quantum correlations effectively identified using inequalities. 
Spin chains, simulated with ultra-cold atoms in optical lattices, Rydberg atoms in tweezer arrays, trapped ions, or molecules, allow single-spin control and measurement. Therefore, they are suitable for studying fundamental aspects of these correlations and non-locality. 
Occupation defects, such as vacancies or multiple atoms occupying a single site due to imperfect system preparation, limit the detection of Bell correlations. 
We study their effects with the help of a simplified toy model parameterised by the probability $p$ of having a single occupation for a given site. Within this model, and for entangled systems obtained by one-axis twisting evolution from an initial factorised state, we derive two Bell inequalities, one based on many-site correlations and the other on two-site correlations, and identify the smallest probability $p$ that allows the Bell inequalities violation to be detected. 
We then consider two physical realizations using entangled ultra-cold atoms in optical lattices where the parameter $p$ is related to a non-unitary filling factor and non-zero temperature. We test the predictions of the toy model against exact numerical results.
\end{abstract}

\maketitle

\def\thefootnote{*}\footnotetext{These authors contributed equally to this work}\def\thefootnote{\arabic{footnote}}

\section{Introduction}

Quantum mechanics introduced groundbreaking concepts of non-local correlations and entanglement that challenged well-established principles of classical physics, including realism, causality and locality~\cite{PhysRev.47.777}. 
A classical picture can be recovered at the price of introducing local hidden variables, and the corresponding local hidden variable (LV) theories are shown to satisfy Bell inequalities~\cite{PhysicsPhysiqueFizika.1.195}. 
Quantum correlations that violate Bell inequalities are inconsistent with LV theory and are, therefore, referred to as non-local~\cite{RevModPhys.86.419}.

The most robust method for entanglement certification is provided by violating Bell's inequalities, as it is independent of assumptions about the physical nature and degrees of freedom to be measured or the calibration of measurements~\cite{PhysRevLett.121.180503}. 
This is known as the device-independent scenario, which is a powerful resource in many quantum information tasks, such as self-testing~\cite{mayers2004selftestingquantumapparatus}, randomness amplification and expansion~\cite{Colbeck2012,Brandao2016}, quantum key distribution~\cite{PhysRevLett.98.230501, roydeloison2024deviceindependentquantumkeydistribution,Nadlinger2022}, and quantum sensing and metrology~\cite{PhysRevA.99.040101,PhysRevLett.126.210506}.
Recent experimental studies of Bell correlations have employed various platforms, including entangled photons~\cite{PhysRevLett.49.1804, PhysRevLett.81.5039,PhysRevLett.115.250401}, spins in nitrogen-vacancy centers~\cite{Hensen2015}, superconducting circuits~\cite{Storz2023}, pairs of Josephson phase qubits~\cite{Ansmann2009}, and ultra-cold neutral atoms~\cite{PhysRevLett.119.010402, doi:10.1126/science.aad8665}.
Many studies have concentrated on few-particle scenarios; however, non-local correlations also naturally emerge in quantum many-body systems~\cite{doi:10.1126/science.1247715}.

Ensembles of atoms in optical lattices and optical tweezer arrays~\cite{Yamamoto_2016, Schine2022,PhysRevX.9.041052}, as well as trapped ions and molecules~\cite{Micheli2006, PRXQuantum.3.030339}, are ideal platforms for studying many-body entanglement, with capabilities for single-atom preparation, control, and detection.
Entanglement can be generated in these platforms by various methods, effectively implementing a coherent evolution that is well approximated by the one-axis twisting (OAT) model~\cite{PhysRevA.47.5138} where spin-squeezed and GHZ states are produced.
Imperfections, however, e.g. in system preparation, may introduce occupation defects, limiting the possibility of detecting Bell's inequalities violation and applying them for entanglement certification.

In this paper, we study the role of occupation defects in detecting Bell correlations focusing on $N$ two-level bosonic atoms in a spin-entangled state distributed among $M$ sites of an optical lattice, with $N\le M$.

\begin{figure}[]
    \centering    \includegraphics[width=\linewidth]{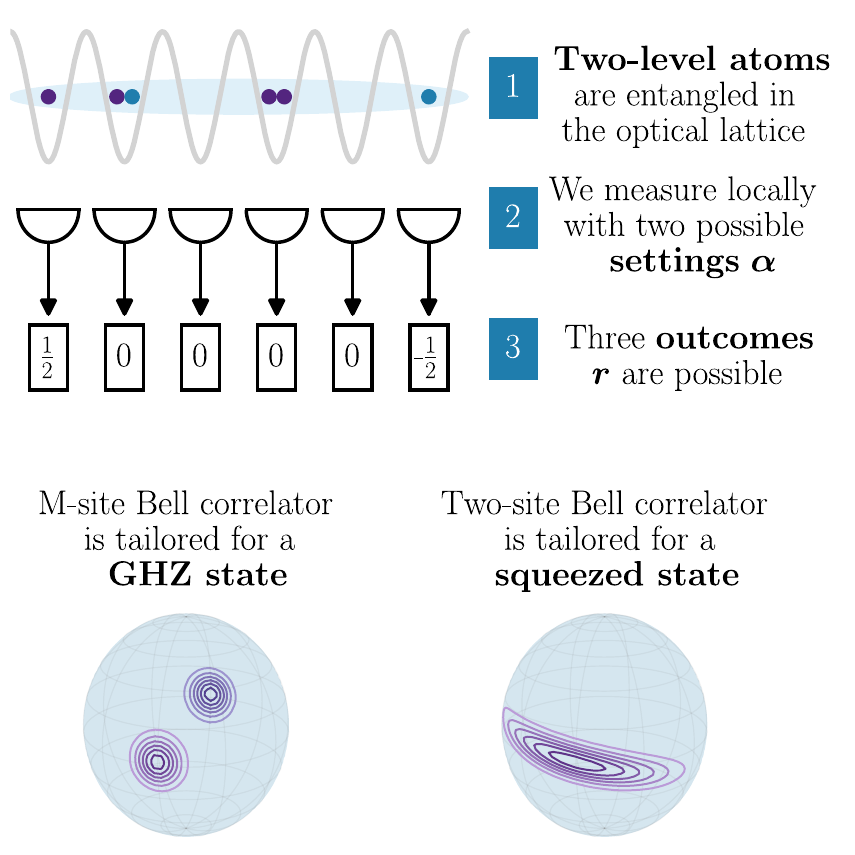}
    \caption{A chain of $M$ spins is created by two-level ultra-cold atoms loaded into an optical lattice. Bell scenario considered: on each of $M$ subsystems (sites) two different measurement are performed $\hat{s}^{(j)}_{\alpha}$, with $\alpha=0,1$, locally leading to three measurement outcomes $r^{(j)}_1,\, r^{(j)}_2, r^{(j)}_3\in\{-1/2,+1/2,0\}$.
    The outcome ``0" is associated to empty or double occupied sites.
   The M-site Bell correlator is tailored for GHZ states, while the two-site Bell correlator is tailored for spin-squeezed states.
   }
    \label{fig:Bell_Scenario}
\end{figure}

In Section~\ref{sec:Bellscenario} we present the Bell scenario in which we consider the measurement of two local collective spin observables at each lattice site where $0$, $1$ or $2$ atoms are present. For each measurement result, there are three possible outcomes: $\pm1/2$ when the site is singly occupied, and $0$ whenever a defect such as an empty or doubly occupied site is present, as illustrated in Fig.~\ref{fig:Bell_Scenario}.

In Section~\ref{sec: ToyModel} we introduce a simplified toy model where the atoms' internal and external degrees of freedom are decoupled. The dynamics of the internal spin degrees of freedom, including possible entanglement among the spins, is assumed to be given by the OAT model with $N$ atoms. The state of the external degrees of freedom is parameterized by the probability $p$ of having a single occupied site allowing vacant, double, etc. occupancy sites when $p<1$. 
We derive the corresponding $p$-parameterized Bell inequalities based on $M$- and two-sites Bell correlations~\cite{Plodzien2022, PRXQuantum.2.030329}. We analytically estimate the lowest (critical) value of the probability $p$ for violation of Bell inequality.
We obtain $p_c= 2^{1/M}/\sqrt{2}$ for the $M$-site Bell correlations measured for a GHZ state. For the two-site Bell correlations measured for a spin-squeezed state, we obtain  $p_c=4/5$ and $p_c=\sqrt{3}/2$ when $N=pM<M$ and $N=M$, respectively. These results are general and may be relevant for any platform where individual addressing of spins is
possible~
\cite{Richerme2014, Kajtoch_2018, PhysRevA.102.013328, PhysRevResearch.1.033075,PhysRevLett.129.150503, PhysRevLett.129.113201,mamaev2023spin, PhysRevA.107.033318}, for example as in recent experiments using an array
of trapped ions~\cite{Franke2023} and Rydberg atoms~\cite{Bornet2023, Eckner2023} demonstrating generation of entanglement in terms of scalable spin-squeezing with tens of spins.

We explore the prediction of the toy model by performing full many-body simulations of ultra-cold atoms in optical lattices described by the Bose-Hubbard defined in Section~\ref{sec:appl}. We consider two distinct protocols for Bell correlations generation, which differ by the source of the occupation defects. 

In the first protocol, discussed in Section~\ref{sec:holes}, Bell correlations between individual atoms are generated in the Mott insulating regime via a weak inhomogeneous magnetic field or interaction anisotropy~\cite{PhysRevB.109.214310,PhysRevLett.129.090403}, and the considered occupation defects are vacant sites (holes). 
The probability $p$ of single occupancy of a given site is here equal to the filling factor $f=N/M$. 
Our numerical simulations confirm predictions of the toy model of the lower bound on the filling factor concerning the detection of the $M$-sites Bell correlations.
Concerning the two-site Bell correlations, we find that their detection is hindered whenever $f<1$ for fixed positions of holes. 
On the contrary, in the case of weak interaction anisotropy, the movement of holes lowers the minimal filling factor allowing for the detection of two-site Bell correlations down to $f\approx p_c$, as predicted by the toy model.

In the second protocol, investigated in Section~\ref{sec: SF}, entanglement is generated in the superfluid regime using atom-atom interactions~\cite{KajtochEPL2018,PhysRevA.102.013328,Plodzien2022} and transferred to the Mott phase by increasing the lattice height. The main source of defects when $N=M$ is imposed by non-zero temperature $T$ of an initial sample. We relate analytically the probability $p$ with the initial temperature $T$ of the system in the limit of an isoentropic transformation that leaves the system in thermal equilibrium at each moment. In this case, we determine analytically the upper bound for the initial temperature corresponding to $p_c$ below which $M$- and two-site Bell correlations can be detected. 

Our results show that for large systems the lower bounds on the probability of defects are weakly dependent on the number of particles demonstrating the possibility of witnessing Bell correlations in lattice systems within the range of typical experimental parameters, as concluded with further discussion in Sec.~\ref{sec:summ}.

\section{Bell Scenario}
\label{sec:Bellscenario}

We consider $M$ spatially separated subsystems labelled $i=1,...,M$, sharing a quantum $M$-partite state described by the density operator $\hat\rho$. In each subsystem, $k$ observables labeled $\alpha=0,...,k-1$, can be measured, each having $d$ possible outcomes. A Bell experiment, as shown in Fig. \ref{fig:Bell_Scenario}, 
involves choosing an arbitrary observable for each of the $M$ subsystems $\alpha=\{\alpha_i\}_{i=1}^M$,
and recording the measurement results $r=\{r^{(i)}\}_{i=1}^M$. The goal is to determine the $M$-partite probability distribution $P_M(r|\alpha)$ for the outcomes $r$ given the settings $\alpha$ for all possible choices of settings $\alpha$. In the LV theory~\cite{Brunner2014}, any probability distribution $P_M(r|\alpha)$ can be written as
\begin{align}\label{eq: DistProbLocal}
P_M^{(\rm LV)}(r|\alpha)=\int d\lambda\,q(\lambda)\prod_{i=1}^MP^{(i)}(r^{(i)}|\alpha_i,\lambda).
\end{align}
The local nature of this equation resides in the fact that the probability distribution of the outcomes for any given subsystem $i$ depends only on the setting within that same subsystem $i$. The correlations between different subsystems here take their origin from a dependence relation that was established in the past, when the state $\hat\rho$ was generated. This dependence can be fully described by some variable $\lambda$, which may be random with a probability distribution $q(\lambda)$, affecting simultaneously all the $M$ subsystems. Furthermore, if we consider a local realistic model where the measurement result $r^{(i)}$ is deterministically determined by the setting $\alpha_i$ and the variable $\lambda$ in each subsystem $i$, we have
\begin{align}\label{eq: DistProbLocalRealistic}
P^{(i)}(r^{(i)}|\alpha_i,\lambda)=\delta_{r^{(i)},r^{(i)}(\lambda)}.
\end{align}

In the Bell experiment described above, the possibility of decomposing the measured probability distribution $P_M(r|\alpha)$ into the form (\ref{eq: DistProbLocal}) constitutes the locality condition for correlations between the $M$ subsystems under study. Conversely, if the measured probability distribution $P_M(r|\alpha)$ cannot be written in the form (\ref{eq: DistProbLocal}),  the correlations present in the system are non-local. In practice, as we will see later, this condition is often formulated as a Bell inequality. 

In this paper, we consider an optical lattice with $M$ sites, each containing zero, one or two two-level atoms. 
Each atom with two internal states $a$ and $b$ is an effective spin $1/2$, and we can express local collective spins in the second quantized form as
$\hat{s}^{(j)}_z=(\hat{a}_j^\dagger\hat{a}_j - \hat{b}_j^\dagger\hat{b}_j)/2$ and
$\hat{s}_x^{(j)} = (\hat{s}_+^{(j)}+\hat{s}_-^{(j)})/2$, 
$\hat{s}_y^{(j)} = (\hat{s}_+^{(j)}-\hat{s}_-^{(j)})/(2i)$, 
with $\hat{s}_+^{(j)} = \hat{a}_j^\dagger\hat{b}_j$, 
$\hat{s}_-^{(j)}=(\hat{s}_+^{(j)})^\dagger$.

Within the Bell scenario, we consider in each site $j$, the measurement of two local collective spin observables $\hat s^{(j)}_{\alpha}$ on the $j$-th site,
with $\alpha=0,1$, each giving $d=3$ possible results
\begin{align}\label{eq: ThreeOutcomes}
r_0^{(j)}=0,-\frac{1}{2},+\frac{1}{2}\quad;\quad r_1^{(j)}=0,-\frac{1}{2},+\frac{1}{2},
\end{align}
as illustrated in Fig.\ref{fig:Bell_Scenario},
where the result $r=0$ is assigned to any measurement result different from $\pm1/2$. The probability distribution $P_M(r|\alpha)$ to obtain the results $r$ given the settings $\alpha$ can be theoretically calculated, in terms of the density operator $\hat\rho$ describing the system's state, as
\begin{align}\label{eq: DistProb}
P_M(r|\alpha)={\rm tr}\left[\hat\rho\bigotimes_{j=1}^M\hat\Pi_{\alpha_j,r^{(j)}}\right]
,\end{align}
where $\hat\Pi_{\alpha_j,r^{(j)}=\pm1/2}$ projects onto the eigensubspace of $\hat s^{(j)}_{\alpha_j}$ with eigenvalue $r^{(j)}=\pm1/2$, and $\hat\Pi_{\alpha_j,r^{(j)}=0}=\hat{\mathbb{1}}-\sum_{r=\pm1/2}\hat\Pi_{\alpha_j,r}$ projects onto the subspace perpendicular to both eigensubspaces of $\hat s^{(j)}_{\alpha_j}$ corresponding to the eigenvalues $r^{(j)}=+1/2$ and $r^{(j)}=-1/2$. 

In the next Section, we derive two Bell inequalities using this framework. The first one, relying on $M$-site correlations, is mainly useful for highly entangled states with a small number of lattice sites (see, e.g., Fig. \ref{fig:Bell_Scenario}). The second one, relying on two-site correlations, is more suitable to spin-squeezed states with a large number of lattice sites (see, e.g., Fig. \ref{fig:Bell_Scenario}). 
For both cases, we introduce a simplified toy model accounting for occupation defects that result in measurement outcomes $r^{(j)}=0$.

\section{Toy model}\label{sec: ToyModel}

In the lattice system considered one starts from a 
product state $|x\rangle^{\otimes N}$ where each atom is in a coherent superposition of two internal states, i.e. a coherent spin state (CSS) along the $x$ direction.
The entanglement between spins is dynamically obtained using many-body interactions by different protocols (see Sec. \ref{sec:holes} and Sec. \ref{sec: SF}) that can be effectively described, in some limit, by unitary evolution with the OAT Hamiltonian
\begin{align}\label{eq: OAT}
\hat H_{\rm OAT} = \hbar\chi \hat S_z^2,
\end{align}
where $\chi$ quantifies the strength of interactions and $\hat S_\sigma$ represents the collective spin operator of $N$ two-level atoms with $\sigma = x,y,z$.
The OAT model generates spin-squeezed states as well as non-Gaussian entangled states, including the GHZ state~\cite{YURKE1988298,PhysRevA.78.023606}.

To study the role of occupation defects in the detection of Bell correlations, we introduce an approximate toy model where the internal and external degrees of freedom of the atoms are decoupled 
\begin{align}\label{eq:system-state}
\hat{\rho} = \hat{\rho}_{\mathrm{ext}}\otimes\hat{\rho}_{\mathrm{SS}}.
\end{align} 
Here $\hat\rho_{\rm SS}$ describes the internal degrees of freedom of the atoms including entanglement among the spins. In particular, we consider the OAT evolution (\ref{eq: OAT})
\begin{align}\label{eq: OAT_dynamics}
\hat{\rho}_{\mathrm{SS}}=|\psi_t\rangle\langle\psi_t|\quad\textrm{with}\quad|\psi_t\rangle=e^{-i\hat H_{\rm OAT}t/\hbar}|x\rangle^{\otimes N},
\end{align}
and $\hat\rho_{\rm ext}$ describes the external degrees of freedom, which we assume to be factorized over the different lattice sites 
\begin{align}\label{eq:extern-state}
\hat{\rho}_{\mathrm{ext}} = \bigotimes_{j=1}^M\hat \rho_j\quad\textrm{with}\quad \hat\rho_j=p |1\rangle_j {}_j\langle 1| +(1-p) \hat\rho^{\perp}_j,
\end{align}
where $|1\rangle_j$ is the Fock state with one spin at site $j$ and $p$ is the probability of having a single occupancy state. The probability of having an empty, double, etc.,  occupied site, as described by $\hat\rho^{\perp}_j$, is given by $(1-p)$. In practice, in $\hat\rho^{\perp}_j$ we only consider empty sites or doubly occupied sites. In this framework, when measuring a local spin observable, we assign the value $r^{(j)}=0$ to any measurement result different from $\pm1/2$. 

\subsection{$M$-site Bell correlations}

We now derive a Bell inequality whose violation allows the detection of non-local correlations. For this, from the measurement outcomes, in the two settings $\alpha=0,1$, in each subsystem we introduce the complex quantity 
\begin{align}
r_+^{(j)} = r_0^{(j)}+i\,r_1^{(j)}
\end{align}
and we consider the product
\begin{align}\label{eq: Msites_corr_fun}
c_M = \prod_{j=1}^M r_+^{(j)},
\end{align} 
In a local realistic theory, where the probability distribution $P_M(r|\alpha)$ has the form of (\ref{eq: DistProbLocal}) and (\ref{eq: DistProbLocalRealistic}), the average of $c_M$ over many Bell experiment realizations is given by 
\begin{align}\label{eq: Avrage}
\langle c_M\rangle=\int d\lambda\,q(\lambda)\prod_{j=1}^M r_+^{(j)}(\lambda).
\end{align}
By introducing the functions $f(\lambda)=1$ and $g(\lambda)=\prod_{j=1}^M r_+^{(j)}(\lambda)$, and the scalar product $\langle f,g\rangle=\int d\lambda\, q(\lambda)f^*(\lambda)g(\lambda)$, the application of the Cauchy-Schwarz inequality to the functions $f(\lambda)$ and $g(\lambda)$, gives
\begin{align}
|\langle f,g\rangle|^2\le\langle f,f\rangle\langle g,g\rangle.
\end{align}
This leads to the following Bell inequality
\begin{align}\label{eq: BellIneq_1}
\mathcal{E}_M\equiv|\langle c_M\rangle|^2\leq\int d\lambda\, q(\lambda)\prod_{j=1}^M|r_+^{(j)}(\lambda)|^2\le 2^{-M},
\end{align}
and $\mathcal{E}_M$ is the $M$-site Bell correlator.
We choose, in each site $j$, the settings $\alpha=0$ and $\alpha=1$ corresponding to $\hat s_y^{(j)}$ and $\hat s_z^{(j)}$ respectively \cite{Plodzien2022}, to form 
\begin{align}\label{eq:settingMsite}
\hat s_+^{(j)}=\hat s_y^{(j)}+i\,\hat s_z^{(j)}.
\end{align}
It is worth stressing here, that the above setting is optimal for the GHZ state and is less effective for other entangled states generated by the OAT dynamics.
The Bell inequality (\ref{eq: BellIneq_1}) takes the form
\begin{align}\label{eq: SpinBell}
\mathcal{E}_M=|\langle\hat s_+^{(1)}\hat s_+^{(2)}...\hat s_+^{(M)}\rangle|^2\le 2^{-M}.
\end{align}

When considering the toy model (\ref{eq:system-state}), the only nonzero contribution to ${\cal E}_M$ (\ref{eq: SpinBell}) comes from the sites occupied by a single atom. We have
\begin{align}
\label{eq: corrTP0}
  \mathcal{E}_{M}^{(p\ne1)} = p^{2M} \mathcal{E}_{M}^{(p=1)},
\end{align}
where $p^M$ represents the probability that the $M$ sites are occupied with a single atom. In the non-Gaussian regime of the OAT dynamics (\ref{eq: OAT_dynamics}), a macroscopic superposition of coherent states is created at some particular instants $\chi \tau=\pi/q$ labelled by an even integer $q=2, 4, 6, …, M$, with $q=2$ for the GHZ state \cite{Plodzien2022,PhysRevA.78.023606}. The $M$-site Bell correlator corresponding to these non-Gaussian spin states, for $p=1$, reads $\mathcal{E}_{M}^{(p=1)}\eqsim 1/q^2$, see Eq.(17) in~\cite{Plodzien2022}. By replacing in (\ref{eq: corrTP0}), one can derive a critical value of $p$ below which the $M$-site Bell correlations present in the states generated at $\chi \tau=\pi/q$ cannot be detected
\begin{align}
\label{eq: p_c}
  p_{\rm c} = \frac{q^{1/M}}{\sqrt 2},
\end{align}
i.e. for which $\mathcal{E}_{M}^{(p=p_{\rm c})} = 2^{-M}$.
We note that in the large $M$ limit we have $p_c\approx 1/\sqrt{2}$ for all $q$.

\subsection{Two-sites Bell correlations}
\label{sec:two-site-bell-general}

In the case of the two-sites Bell correlations, we introduce the vector $\vec M$ and the matrix $\tilde{C}$, whose elements, for $\alpha,\beta=0,1$, are respectively given by
\begin{align}\label{eq: Average}
M_\alpha &= \sum_{j=1}^M\langle r^{(j)}_{\alpha}\rangle\\ \label{eq: Correlations}
C_{\alpha \beta}&=\sum_{i,j\neq i}\langle r^{(i)}_{\alpha}r^{(j)}_{\beta}\rangle, \quad \tilde{C}_{\alpha \beta} = C_{\alpha \beta}-M_\alpha M_\beta .
\end{align}
It can be shown \cite{PRXQuantum.2.030329} that for any input data $\vec{M}$ and $\tilde{C}$ compatible with a LV theory (\ref{eq: DistProbLocal}), any $2\times 2$ positive semi-definite matrix $A$ and any $2\times 1$ vector $\vec{h}$, the following Bell inequality holds
\begin{align}\label{eq:GeneralBell}
L(A,h) = \mathrm{tr}(A\tilde{C})+\vec{h}\cdot\vec{M} + E_{\rm max}\geq 0,
\end{align} 
where the classical limit is
\begin{align}
E_{\rm max}=\max_{\vec{r}}\left[\vec{r}\,^TA\vec{r}-\vec h\cdot\vec{r}\right],
\end{align}
with $\vec{r}=(r_0,r_1)^T$ being the vector of all possible pairs of outcomes corresponding to the two settings $\alpha=0$ and $\alpha=1$ for a single subsystem. The positive semi-definite matrix $A$ and the vector $\vec h$ that minimize $L(A,h)$ can be found using a data-driven method as in Ref. \cite{PRXQuantum.2.030329}. We find that the Bell inequality
~\footnote{Alternatively, the Bell inequality (\ref{eq:FirstBell}) can be represented using the $C$ matrix in (\ref{eq: Correlations}), namely
$L=C_{00}+C_{11}-C_{01}-C_{10}
-(M_0-M_1)^2-M_0-M_1 + M\geq 0
$.}
\begin{align}\label{eq:FirstBell}
L=\tilde{C}_{00}+\tilde{C}_{11}-\tilde{C}_{01}-\tilde{C}_{10}-M_0-M_1 + M\geq 0,
\end{align}
first established in Ref. \cite{PRXQuantum.2.030329} for spin-squeezed states in the case of only two possible measurement outcomes, is optimal also in our case with the three possible outcomes (\ref{eq: ThreeOutcomes}) and under occupation defects.

For the case of two-sites Bell correlations, we choose $\alpha=0$ and $\alpha=1$ corresponding to the measurement, at each site $j$, of the spin components \cite{PRXQuantum.2.030329}
\begin{align}\label{eq:settings}
\hat{s}^{(j)}_0 &=  \hat{s}^{(j)}_{\vec n}\cos(\theta)+\hat{s}^{(j)}_{\vec m}\sin(\theta),\\
\label{eq:settings1}
\hat{s}^{(j)}_1 &=  \hat{s}^{(j)}_{\vec n}\cos(\theta)-\hat{s}^{(j)}_{\vec m}\sin(\theta),
\end{align}
where the unit vector $\vec n$ is in the spin direction and $\vec m$, perpendicular to the spin direction, is in the best squeezing direction. The choice of the two settings (\ref{eq:settings})-(\ref{eq:settings1}) is dictated by the geometry of the spin-squeezed states. Non-zero expectation values come from the averages calculated in the plane spanned by the $\vec{n}$ and $\vec{m}$ vectors.

The averages (\ref{eq: Average}) and the correlations (\ref{eq: Correlations}) for $\alpha,\beta\in\{0,1\}$, that form the Bell inequality (\ref{eq:FirstBell}), take respectively the form
\begin{align}\label{eq:AverageSpin}
M_{\alpha} &= \sum_{j=1}^M\sum_{r=\pm1/2}r\,{\rm tr}\left[\hat\rho\,\hat\Pi_{\alpha_j,r}\right] \\ \label{eq:CorrelationsSpin}
\tilde{C}_{\alpha\beta} &=\sum_{i,j\ne i}^M\left(\sum_{r,s=\pm1/2}rs\,{\rm tr}\left[\hat\rho\,\hat\Pi_{\alpha_i,r}\otimes\hat\Pi_{\beta_j,s}\right]\right)-M_{\alpha}M_{\beta}.
\end{align} 

We now evaluate $M_{\alpha}$ and $\tilde{C}_{\alpha\beta}$ using the toy model density matrix (\ref{eq:system-state}), for the physical situation presented in Sec.~\ref{sec:holes} where $N\le M$, and that in Sec.\ref{sec: SF} where $N=M$.

\subsubsection{Non-unit filling}

The first scenario, relevant to the system described in Sec. \ref{sec:holes}, assumes that $N=pM\le M$ where $N$ is the number of atoms and $M$ is the number of sites.
Under this condition, $p<1$ indicates a non-unit filling of the lattice whose sites are empty or singly occupied. The density matrix $\hat\rho^{\perp}_j$ in (\ref{eq:extern-state}), in this scenario, becomes
\begin{align}
\hat\rho^{\perp}_j=|0\rangle_j {}_j\langle 0|.
\end{align}
The averages (\ref{eq: Average}) and the correlations (\ref{eq: Correlations}) for $\alpha,\beta\in\{0,1\}$ are in this case
\begin{align}
\label{eq:average_MI}
M_\alpha &= \langle\hat S_{\vec{n}}\rangle_{\rm SS}\cos(\theta),\\ 
\label{eq:corr_MI}
\tilde{C}_{\alpha \beta} &=\frac{N-p}{N-1}
\left\{
(\Delta\hat S_{\vec{n}})^2_{\rm SS}\cos^2(\theta)-(-1)^{\delta_{\alpha \beta}}(\Delta\hat{S}_{\vec m})^2_{\rm SS}\sin^2(\theta) \right. \nonumber\\
{}& \left. 
-\frac{N}{4}
\cos[ 2 \theta (1-\delta_{\alpha \beta}) ]
\right\}
+\frac{1-p}{N-1}\langle\hat S_{\vec n}\rangle^2\cos^2(\theta).
\end{align} 
where the subscript ${\rm SS}$ refers to an expectation calculated in the $\hat \rho_{\rm SS}$ state (\ref{eq: OAT_dynamics}).

\begin{figure}[t]
    \centering
    \includegraphics[width=\linewidth]{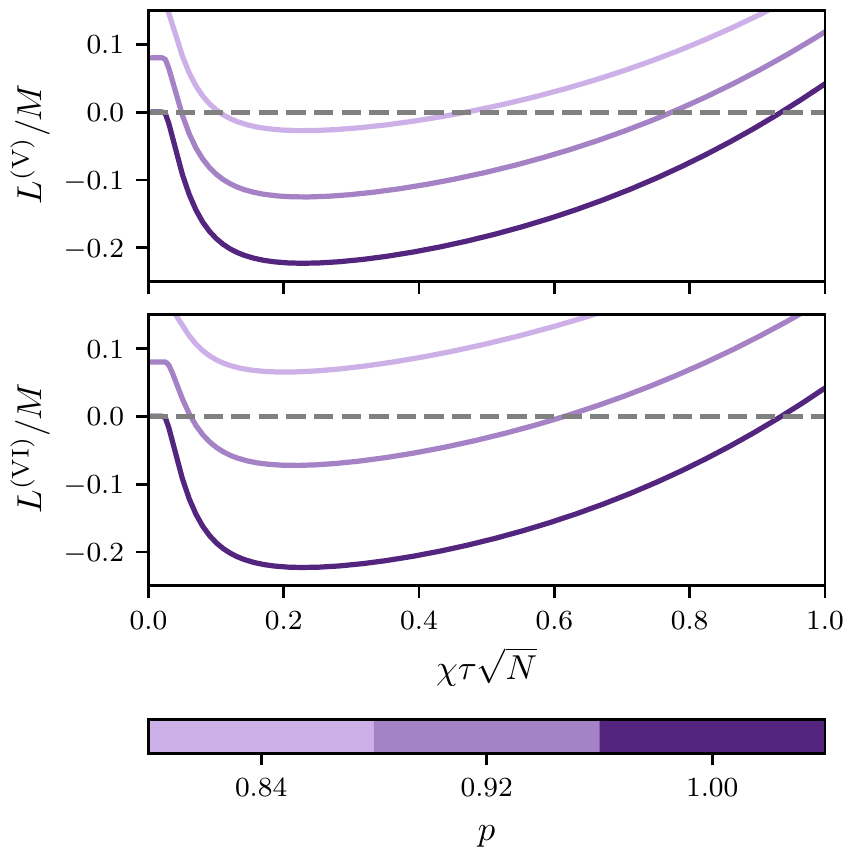}
    \caption{Variation of the non-locality witness $L_{\rm opt}/M$, given by  (\ref{eq:L_opt_pM}) in the upper panel and (\ref{eq:L_opt}) in the lower panel, during the one-axis twisting dynamics with $N=10^3$ for different values of $p$
    (probability of single occupation of a given site).
    The flat behaviour at short times is a consequence of the optimization condition.
    }
    \label{fig:NL_correlations}
\end{figure}

The Bell inequality (\ref{eq:FirstBell}) is given, for this scenario, by
\begin{align}\label{eq:Bell_Inequality}\nonumber
\frac{L^{\rm (V)}_{\theta}}{M}&=\frac{p(N-p)}{N-1}\sin^2(\theta)\left(\frac{4(\Delta\hat{S}_{\vec m})^2_{\rm SS}}{N}-1\right)\\
&\quad-2p\frac{\langle\hat{S}_{\vec n}\rangle_{\rm SS}}{N}\cos(\theta)+1\geq0.
\end{align}
The minimization of $L^{\rm (V)}_{\theta}$ with respect to $\theta$ gives
\begin{align}\label{eq:L_opt_pM}\nonumber
\frac{L^{\rm (V)}}{M}&=1-\frac{p(N-1)}{N-p}\frac{\langle\hat{S}_{\vec n}\rangle^2_{\rm SS}}{N\left[N-4(\Delta\hat{S}_{\vec m})^2_{\rm SS}\right]}\\
&\quad-\frac{p(N-p)}{N-1}\left[1-\frac{4(\Delta\hat{S}_{\vec m})^2_{\rm SS}}{N}\right],
\end{align}
for an optimal $\theta$ such that $\cos\theta_{\rm opt}=\frac{N-1}{N-p}\frac{\langle\hat{S}_{\vec n}\rangle_{\rm SS}}{N-4(\Delta\hat{S}_{\vec m})^2_{\rm SS}}$\footnote{This solution should verify $|\cos(\theta_{\rm opt})|\leq1$, which introduces the restriction $p\le N-(N-1)|\langle\hat S_{\vec n}\rangle\big{/}[N-4(\Delta\hat S_{\vec m})^2]|$. Here, $L^{\rm (V)}$ can be written in terms of the squeezing parameter $\xi^2=N (\Delta\hat{S}_{\vec m})^2_{\rm SS}\big{/}\langle\hat{S}_{\vec n}\rangle_{\rm SS}^2$  and the normalized mean spin $v=\langle\hat{S}_{\vec n}\rangle_{\rm SS}\big{/}(N/2)$ as
\begin{equation}
    \frac{L^{\rm (V)}}{M}= 1-\frac{p(N-1)}{N-p}\frac{v^2/4}{1-\xi^2 v^2}-\frac{p(N-p)}{N-1}(1-\xi^2v^2).
\end{equation}
}. 
The equation (\ref{eq:L_opt_pM}) is represented in Fig. \ref{fig:NL_correlations} (upper panel) as a function of time of the OAT dynamics for a given $N$ and for different values of the occupation probability $p$. This reveals that for a given $N$ and at each moment $\tau$, there is a critical value of $p$ below which, the two-site Bell correlations present in the system cannot be detected. This critical probability $p=p_{\rm c}$ is the value for which, the following $3^{\rm rd}$ order equation holds
\begin{align}\nonumber
&p^3-2N\left(1-\frac{4(\Delta\hat{S}_{\vec m})^2_{\rm SS}}{N}\right)p^2
+\left[N^2 + N-1 \right] p \nonumber \\
&+\left[4N (\Delta\hat{S}_{\vec m})^2_{\rm SS}+\frac{(N-1)^2\langle\hat{S}_{\vec n}\rangle^2_{\rm SS}}{N\left[N-4(\Delta\hat{S}_{\vec m})^2_{\rm SS}\right]}\right]p \nonumber \\
&-N(N-1)=0.
\end{align}
In the limit of large atom number $N$, the minimal value of (\ref{eq:L_opt_pM}), over $\chi \tau$, converges to the $p$-dependent constant 
\begin{align}\label{eq: L_opt_limit_Tana}
\frac{L^{\rm (V)}_{\rm min}}{M}\approx 1-\frac{5}{4}p,
\end{align}
and the corresponding critical value of the occupation probability tends to $p_{\rm c}=4/5$.

\subsubsection{Double occupancies and empty sites}

In a second scenario, relevant to the system described in Sec. \ref{sec: SF}, we assume that $N=M$ where $N$ is the number of atoms and $M$ is the number of sites. Under this condition, $p<1$ indicates a redistribution of the $N$ atoms among the $M$ lattice sites resulting in the emergence of both empty and doubly occupied sites. The filling factor is one, $f=1$. Thus, the density matrix $\hat\rho^{\perp}_j$ in (\ref{eq:extern-state}) can be written as
\begin{align}
\hat\rho^{\perp}_j=\frac{|0\rangle_j {}_j\langle 0|+|2\rangle_j {}_j\langle 2|}{2}.
\end{align}
The calculation of the averages (\ref{eq: Average}) and the correlations~(\ref{eq: Correlations}) for $\alpha,\beta\in\{0,1\}$ gives
\begin{align}\label{eq:averagee}
M_\alpha &= p \langle\hat S_{\vec{n}}\rangle_{\rm SS}\cos(\theta),\\ \label{eq:corr}\nonumber
\tilde{C}_{\alpha \beta} &=p^2 (\Delta\hat S_{\vec{n}})^2_{\rm SS}\cos^2(\theta)-(-1)^{\delta_{\alpha \beta}}p^2 (\Delta\hat{S}_{\vec m})^2_{\rm SS}\sin^2(\theta)\\
&\quad-p^2\frac{N}{4}\cos[2\theta(1-\delta_{\alpha \beta})], 
\end{align} 
where the subscript ${\rm SS}$ refers to an expectation calculated in the $\hat \rho_{\rm SS}$ state (\ref{eq: OAT_dynamics}), and the Bell inequality (\ref{eq:FirstBell}) reads
\begin{align}\label{eq:LAh}
\frac{L_{\theta}^{\rm (VI)}}{M}&=p^2\sin^2(\theta)\left(\frac{4(\Delta\hat{S}_{\vec m})^2_{\rm SS}}{N}-1\right) \nonumber \\
&-2p\frac{\langle\hat{S}_{\vec n}\rangle_{\rm SS}}{N}\cos(\theta)+1\geq0.
\end{align}
By minimizing $L^{\rm (VI)}_{\theta}$ with respect to $\theta$ we obtain
\begin{align}\label{eq:L_opt}
\frac{L^{\rm (VI)}}{M}=1-\frac{\langle\hat{S}_{\vec n}\rangle^2_{\rm SS}}{N\left[N-4(\Delta\hat{S}_{\vec m})^2_{\rm SS}\right]}-p^2\left[1-\frac{4(\Delta\hat{S}_{\vec m})^2_{\rm SS}}{N}\right],
\end{align}
for an optimal $\theta$ such that $\cos\theta_{\rm opt}=\frac{1}{p}\frac{\langle\hat{S}_{\vec n}\rangle_{\rm SS}}{N-4(\Delta\hat{S}_{\vec m})^2_{\rm SS}}$\footnote{This solution should verify $|\cos(\theta_{\rm opt})|\leq1$ which introduces the restriction $p\ge |\langle\hat S_{\vec n}\rangle\big{/}[N-4(\Delta\hat S_{\vec m})^2]|$.}\textsuperscript{,}\footnote{An alternative formula of $L^{\rm (VI)}$ can be obtained by introducing the spin squeezing parameter $\xi^2$ and the normalized mean spin $v$ as
\begin{equation}
    \frac{L^{\rm (VI)}}{M}= 1-\frac{v^2/4}{1-\xi^2 v^2} - p^2(1-\xi^2v^2).
\end{equation}
}. 
The equation (\ref{eq:L_opt}) is represented in Fig. \ref{fig:NL_correlations} (lower panel) as a function of time $\tau$ of the OAT dynamics for a given $N$ and for different values of $p$. We note that for a given $N$ and at each moment, there is a critical value of $p$ below which, the two-site Bell correlations present in the system cannot be detected, from (\ref{eq:L_opt}) we obtain
\begin{align}\label{eq:pcr2data}
p_{\rm c} = \frac{\sqrt{N\left[N-4(\Delta\hat{S}_{\vec m})^2_{\rm SS}\right]-\langle\hat{S}_{\vec n}\rangle^2_{\rm SS}}}{N-4(\Delta\hat{S}_{\vec m})^2_{\rm SS}}.
\end{align}
In the limit of large atom number $N$, the minimal value of (\ref{eq:L_opt}), over $\chi t$, approaches a $p$-dependent constant value
\begin{align}\label{eq: L_opt_limit}
\frac{L^{\rm (VI)}_{\rm min}}{M}\approx \frac{3}{4}-p^2,
\end{align}
and the corresponding critical probability $p_{\rm c}$ approaches $p_{\rm c}=\sqrt{3}/2$.  A lower value of $p$ can be obtained with a Bell inequality including onsite and two-site correlations up to the fourth order. In the limit of large $N$, this leads to a critical occupation probability of $p_{\rm c}=1/2$ \cite{BaamaraThese2023}.

\section{Application with ultra-cold atoms in optical lattices} 
\label{sec:appl}

We test the predictions of our toy model using a system consisting of $N$ ultra-cold bosonic atoms confined in an optical lattice. We focus on rubidium-87 atoms occupying two internal states, labelled $a$ and $b$, and loaded into an optical lattice potential, akin to recent experiments employing quantum gas microscopes~\cite{PhysRevX.9.041014, doi:10.1126/science.abk2397}. The optical lattice comprises $M$ lattice sites and is considered to be in one dimension.

Using the Wannier functions basis~\cite{RevModPhys.80.885}, when the system is in the lowest energy band and 
in the tight-binding limit, the lattice potential exceeding the recoil energy $E_R=\hbar^2k^2/(2m)$, the dynamics is conveniently described by the Bose-Hubbard Hamiltonian
\begin{align}\label{eq:BHM}
\hat{\mathcal{H}}_{\rm BH} &= - t \sum\limits_{i, j=i\pm 1} \left(\hat{a}_{i}^{\dagger}\hat{a}_{j} + \hat{b}_{i}^{\dagger}\hat{b}_{j}\right) + \frac{U_{aa}}{2}\sum\limits_{i=1}^M \hat{n}^a_i (\hat{n}^a_i -1) \nonumber\\
& + \frac{U_{bb}}{2}\sum\limits_{i=1}^M \hat{n}^b_i (\hat{n}^b_i -1) 
+ U_{ab} \sum \limits_{i=1}^M \hat{n}^a_i\hat{n}^b_i ,
\end{align}
where $t$ and $U_{\sigma \sigma'}$ are the tunneling and interaction parameters.
$\hat{a}_i$ ($\hat{b}_i$) is the annihilation operator of an atom in internal state $a$ ($b$) in the $i$-th site of the lattice, and $\hat{n}^a_i=\hat{a}_{i}^{\dagger}\hat{a}_{i}$, $\hat{n}^b_i=\hat{b}_{i}^{\dagger}\hat{b}_{i}$ are the corresponding number operators. 

We explore two protocols for the dynamical generation of Bell correlations within this system as detailed in Sections~\ref{sec:holes} and~\ref{sec: SF}, both resulting in Mott entangled states where atoms exhibit spin entanglement and are distributed across the lattice. 
In both cases, the initial state $\hat{\rho}_a$ with all atoms in the internal state $|a\rangle$ is turned into a
coherent superposition of $|a\rangle$ and $|b\rangle$ by applying a $\pi/2-$pulse, which is equivalent to 
\begin{equation}
\label{eq:coherentstate}
    \hat{\rho}_{\rm ini} = e^{-i \hat{S}_y \pi/2 } \hat{\rho}_a e^{i \hat{S}_y \pi/2 } .
\end{equation}
The two protocols differ in their specific mechanisms for inducing entanglement in the system and hence in the relevant source of occupation defects.

In the first protocol, Bell correlations are generated directly in the strongly interacting Mott regime $U_{\sigma \sigma'}\gg t$ where individual spins interact via spin-exchange interactions between neighbouring spins~(\ref{eq:XXZ}). 
These interactions allow for the generation of entanglement within the system either through weak anisotropy or by coupling with an inhomogeneous field.
The mechanism for generating entanglement was explained and demonstrated in prior works such as~\cite{PhysRevLett.129.090403, PhysRevB.108.104301, PhysRevB.109.214310}.
In the second protocol, Bell correlations are induced by atom-atom interactions in the superfluid regime $t\gg U_{\sigma \sigma'}$ with all-to-all individual spin connections and transferred to the Mott phase by an adiabatic increase of the optical lattice depth, a process described in detail in previous studies~\cite{Kajtoch_2018,PhysRevA.102.013328, Plodzien2022}.

The final state of both protocols allows measuring spin components at a specific lattice site. In the Bell scenario, the outcomes of local measurements across all lattice sites can be collected, and used for the detection of Bell correlations using inequalities (\ref{eq: SpinBell}) and (\ref{eq:FirstBell}).
In the ideal realization of the Mott phase protocol, when $\hat{\rho}_a$ is the ground state where each lattice site hosts precisely one atom, two distinct measurement outcomes are possible.
However, imperfections introduce non-unit filling throughout the lattice, leading to additional measurement outcomes, such as when a site is either empty or double occupied. These failures arise from imperfect preparation of initial state $\hat{\rho}_a$ or to non-zero temperature.

In the following sections, we consider the role of imperfections and we establish a connection between the factor $p$ of the toy model (\ref{eq:system-state}) and the filling factor $f=N/M$ or the initial temperature $T$ for these two protocols.

\section{Entanglement generation in the Mott phase}
\label{sec:holes}

In the strongly interacting limit $U_{\sigma, \sigma'}\gg t$ and in the ground state manifold with at most one atom per lattice site, the system (\ref{eq:BHM}) is approximately described by the $t$--$J$ model 
\begin{align}
\label{eq:t-J}
	&\hat{H}_{t-J}=- t \sum\limits_{i, j=i\pm 1} \hat{P}_0
 \left(\hat{a}_{i}^{\dagger}\hat{a}_{j} + \hat{b}_{i}^{\dagger}\hat{b}_{j}\right)\hat{P}_0
    + \hat{H}_{\rm XXZ} ,
\end{align}
where $\hat{P}_0$ is a projector operator over the manifold of at most single occupancy states, and where
\begin{equation}
	\hat{H}_{\rm XXZ}=
    - J \sum_{j=1}^{M-1} 
    \bigg( \hat{s}_x^{(j)} \hat{s}_x^{(j+1)} + \hat{s}_y^{(j)} \hat{s}_y^{(j+1)} + \Delta \hat{s}_z^{(j)} \hat{s}_z^{(j+1)} - \frac{1}{4} \bigg),
\label{eq:XXZ}
\end{equation}
is the Heisenberg XXZ model with the spin-exchange interaction parameter $J=4 t^2/U_{ab}$ and the anisotropy parameter $\Delta= U_{ab}/U_{aa} + U_{ab}/U_{bb} - 1$~\cite{Altman_2003}. 
When $\Delta=1$ the Hamiltonian takes the form of the isotropic Heisenberg XXX model and it is a natural case for rubidium-87 where $U_{aa}\approx U_{bb}\approx U_{ab}$. The anisotropy parameter $\Delta$ can be tuned by changing the values of interaction strengths using either Feshbach resonances or by shifting optical lattice potentials for states $a$ and $b$.
The collective spin operators are just a summation over the individual ones,
$\hat{S}_\sigma = \sum_{j=1}^M \hat{s}^{(j)}_\sigma$ for $\sigma = x,y,z, \pm$.
The tunnelling term in (\ref{eq:t-J}) is relevant whenever the filling factor $f=N/M$ is not one, meaning there are holes (empty sites) in the system and $N\le M$. 

Let us consider the case of zero temperature when the initial state $\hat{\rho}_a$ is in the Mott regime. 
In the ideal case, we have $\hat{\rho}_a = \bigotimes_{j=1}^M |a\rangle \langle a|_j$, indicating that at each lattice site there is an atom in the internal state $|a\rangle$. However, in the presence of holes, the on-site state can be $|0\rangle \langle 0|_j$ if it is not occupied. In a given experimental realisation, the number of holes can be arbitrary as well as their positions.
In our simulations, we evolve single realizations with initially each site being empty or singly occupied, and then average over many realizations to evaluate expectation values. Specifically,
 we generate a random number $x_j\in (0,1]$ for each lattice site, obtaining
\begin{equation}\label{eq:psi_delta}
    |\Psi_x\rangle_j = \bar{\theta}(f-x_j) |a\rangle_j 
    + \left( 1-\bar{\theta}(f-x_j) \right) |0\rangle_j,
\end{equation}
where $f$ is the filling factor and $\bar{\theta}(x)$ is the Heaviside step function.
The density matrix $\hat{\rho}_a$ describing $N_r$ realizations is obtained through a simple averaging,
\begin{equation}
    \hat{\rho}_a=
    \frac{1}{N_r} 
    \sum_{l =1}^{N_r} 
    \left[ \bigotimes_{j=1}^M 
    |\Psi_x\rangle_j{}_j \langle \Psi_x |
    \right],
\end{equation}
over the set of random numbers $x=\{x_1^{l}, x_2^{l},  \cdots, x_M^{{l}}\}$ with $l\in \{1,...N_r\}$.
One can show, that in the continuous limit, we obtain
\begin{equation}
\label{app:rho_tot}
    \hat{\rho}_a = \bigotimes_{j=1}^M \left( f |a\rangle_j{}_j  \langle a| + (1-f) |0\rangle  _j {}_j \langle 0| \right),
\end{equation}
given that all the sites are equivalent.
Therefore, the filling factor $f$ can be identified with the parameter $p$ of the external state of the toy model (\ref{eq:system-state}) provided that $\hat{\rho}_j^{\perp}=|0\rangle _j {}_j \langle 0|$. 
Owing to the presence of holes, the average number of atoms $N$ differs from the number of lattice sites $M$ when $f\ne 1$, namely $N=f M$.

It is worth noting here, that the density matrix $\rho_a$ in~(\ref{app:rho_tot}) can also be cast in the following way 
\begin{equation}
    \label{eq:longrhoa}
\begin{split}
    \hat{\rho}_a& = 
    f^M 
    \bigotimes_{j=1}^{M}| a \rangle  _j {}_j\langle a | \\
    &+f^M \left( \frac{1-f}{f}\right)^{1} \left( 
    |0 \rangle  _1 {}_1\langle 0 | \bigotimes_{j=1}^{M-1} | a \rangle  _j {}_j \langle a | + \mathcal{P}_{1} \right) \\
    &+f^M \left( \frac{1-f}{f}\right)^{2} \left( 
    |0 \rangle  _1 {}_1 \langle 0 | \otimes
    |0 \rangle  _2 {}_2 \langle 0 | \bigotimes_{j=1}^{M-2} | a \rangle  _j {}_j \langle a | + \mathcal{P}_{2} \right) \\
    &+ \cdots,
\end{split}
\end{equation}
where $\mathcal{P}_\sigma$ are all other configurations with $\sigma$ holes.
The state (\ref{eq:longrhoa}) can be written using the shorter notation
\begin{equation}
    \hat{\rho}_a = 
    f^M \left[
    \bigotimes_{j=1}^{M} | a \rangle  _j {}_j\langle a | + \sum_{
    \sigma=1}^{M} \left( \frac{1-f}{f}\right)^{\sigma} \Sigma_\sigma\left[  \hat{\rho}_\sigma \right] \right],
    \label{eq:rhoasame}
\end{equation}
where $\hat{\rho}_\sigma$ is the product state of a given configuration having $\sigma$ holes and $1-\sigma$ occupied sites while $\Sigma_\sigma$ represents summation over permutations of all possible configurations.

Having the above expression, the initial spin coherent state (\ref{eq:coherentstate}) can be expressed as follows:
\begin{equation}
\label{eq:coherentstate-holes}
    \hat{\rho}_{\rm ini} =
    f^M \left[ 
    \hat{\rho}_{\rm ini}^{(0)}
    +
    \sum_{
    \sigma=1}^{M} \left( \frac{1-f}{f}\right)^{\sigma} \Sigma_\sigma\left[  \hat{\rho}_{\rm ini}^{(\sigma)} \right] \right],
\end{equation}
where the rotations act on the density matrix corresponding to specific configurations of $\sigma$ holes
\begin{equation}
    \hat{\rho}_{\rm ini}^{(\sigma)} =    
     e^{-i \hat{S}_y \pi/2 } \hat{\rho}_\sigma e^{i \hat{S}_y \pi/2 },
\end{equation}
and $\hat{\rho}_{\rm ini}^{(0)}=e^{-i \hat{S}_y \pi/2 }  \left( \bigotimes_{j=1}^{M} | a \rangle  _j {}_j \langle a | \right) e^{i \hat{S}_y \pi/2 }$.
Therefore, any expectation values of the operator $\hat{O}$ can be evaluated as 
\begin{equation}
    \langle \hat{O} \rangle = f^M \left[ 
    {\rm Tr}[\hat{\rho}_{\rm ini}^{(0)} \hat{O}] + 
    \sum_{\sigma=1}^M 
    \left( \frac{1-f}{f}\right)^{\sigma}
      \langle \hat{O} \rangle_{\sigma} \right],
      \label{eq:expectationholes}
\end{equation}
where $\langle \hat{O} \rangle_{\sigma} = \Sigma_\sigma {\rm Tr}[\hat{\rho}_{\rm ini}^{(\sigma)}\hat{O}] $,
and where summation runs over all configurations of $\sigma$ holes on $M$ lattice sites. 

In Subsections~\ref{subsec:mb-holes} and \ref{subsec:2b-holes}, we employ the above-discussed approach for the numerical evaluation of Bell correlations and the critical value of the filling factor allowing for their detection. 
Before, however, let us discuss the mechanism responsible for generating entanglement within the system in the Mott phase.

\subsection{Effective microscopic description}

To generate entanglement from the initial spin coherent states (\ref{eq:coherentstate-holes}) driven by the $t$--$J$ Hamiltonian (\ref{eq:t-J}) we consider two methods.
Both of them were thoroughly investigated in~\cite{PhysRevB.109.214310}.

The first (A) uses a weak anisotropy $\Delta\ne 1$ in (\ref{eq:XXZ}), such that $\Delta \ll 3 - 2 \cos(\pi/M)$ and $\Delta\gg 2\cos(\pi/M) - 1$, to generate entanglement in the system.
Under an ideal scenario with $N=M$, the system can be represented by a single spin chain whose dynamics is described by the OAT model~\cite{PhysRevB.109.214310}.
The presence of holes divides the spin chain into partial chains separated by holes whose configurations are included in~(\ref{eq:coherentstate-holes}). For instance, if a single hole is located somewhere in the middle of the chain (not at the borders), the system can be effectively viewed as two partial chains. The system dynamics composed of the partial chains separate only when the positions of holes are fixed. 
The microscopic model describing the dynamics of each specific configuration involved in (\ref{eq:longrhoa}) is effectively approximated by the OAT model when the positions of holes are pinned to the lattice. 
The model describing each of $n$ partial chains for a given configuration of $\sigma$ holes reads
\begin{equation}
    \hat{H}^{(0)}_{{\rm eff}, n} = -\chi_{n}^{(0)}  \hat{S}_{z,n}^2 , \,\,\,{\rm with} \,\,\, \chi_{n}^{(0)}=\frac{J (\Delta-1)}{(L_n-1)},
    \label{eq:chi0}
\end{equation}
where $L_n$ is the number of atoms composing the partial chain.

The second method (B) uses a weak inhomogeneous magnetic field 
\begin{equation}
    \hat{H}_B=\sum_j \beta_j \hat{s}^{(j)}_z,
\end{equation}
with $\beta_j \ll J \left[\cos (\frac{\pi}{M} ) - 1\right]$, 
for the isotropic case, $\Delta=1$ to generate entanglement in the system. 
In this case, when $\sigma$ holes are pinned into the lattice, the effective model describing each of the $n$ partial chains is described effectively by the following Hamiltonian
\begin{equation}\label{eq:eff_inhomo}
	\hat{H}_{{\rm eff}, n} = \chi_{ n} 
  \hat{S}_{z,n}^2 + v_{n} \hat{S}_{z,n},
\end{equation}
where we omitted constant energy terms, and where
\begin{align}
    \label{eq:chi1}
	\chi_{n}  &= \frac{1}{L_n-1} \sum_{q=1}^{L_n-1} \frac{|c^{(q)}_n|^2}{E^{(q)}_n},\\    v_n &= \frac{1}{L_n}\sum_{l=l_n}^{l_n+L_n-1} \beta_l,\\
    c^{(q)}_{n}&= \frac{\sqrt{2}}{L_n}\sum_{l=l_n}^{l_n+L_n-1} p_{l}^{(q,n)} (\beta_{l}-v_n),
\end{align}
with $l_n$ being the location of the first spin in the partial chain and $E^{(q)}_n = J (1-\cos(\pi q / L_n))$. 

The effective models for the (A) and (B) methods are derived and explained in detail in~\cite{PhysRevB.109.214310}. Their validity for describing spin-squeezing generation was also demonstrated.

Therefore, one can employ the effective models based on the OAT dynamics for partial chains for evaluating expectation values as given by (\ref{eq:expectationholes}) when considering fixed positions of holes (effectively $t=0$ in the $t$--$J$ model (\ref{eq:t-J})). As we show it in~\ref{subsec:2b-holes}, this model provides an upper bound on the filling factor required for the Bell correlations detection.
The model, however, becomes inaccurate when the tunnelling of holes occurs. In such a case, to test the validity of the toy model, we perform full numerical many-body calculations by the exact diagonalization method of the $t$--$J$ model (\ref{eq:t-J}).

In the next subsections, we demonstrate the generation of Bell correlations in the system using scenarios A and B.
In our numerical simulations, we consider open boundary conditions~\cite{PhysRevB.109.214310} and use the parameters as in the recent experiment of A. Rubio-Abadal et al \cite{PhysRevX.9.041014} with $^{87}$Rb atoms, lattice spacing $d=532$ nm, and inter- and intraspecies interactions $U_{aa} \sim U_{bb} \sim U_{ab}=U$~\cite{Fukuhara2013} with $U=24.4t$.

\subsection{$M$-site Bell correlations}
\label{subsec:mb-holes}

We numerically evaluate the many-body Bell correlator 
$\mathcal{E}_M = |\langle \hat{s}_{1,+} \ldots \hat{s}_{M,+} \rangle |^2$ as defined in (\ref{eq: SpinBell}) for the initial spin coherent state given by (\ref{eq:coherentstate-holes}).

\begin{figure}[]
    \centering
    \includegraphics[width=1.\linewidth]{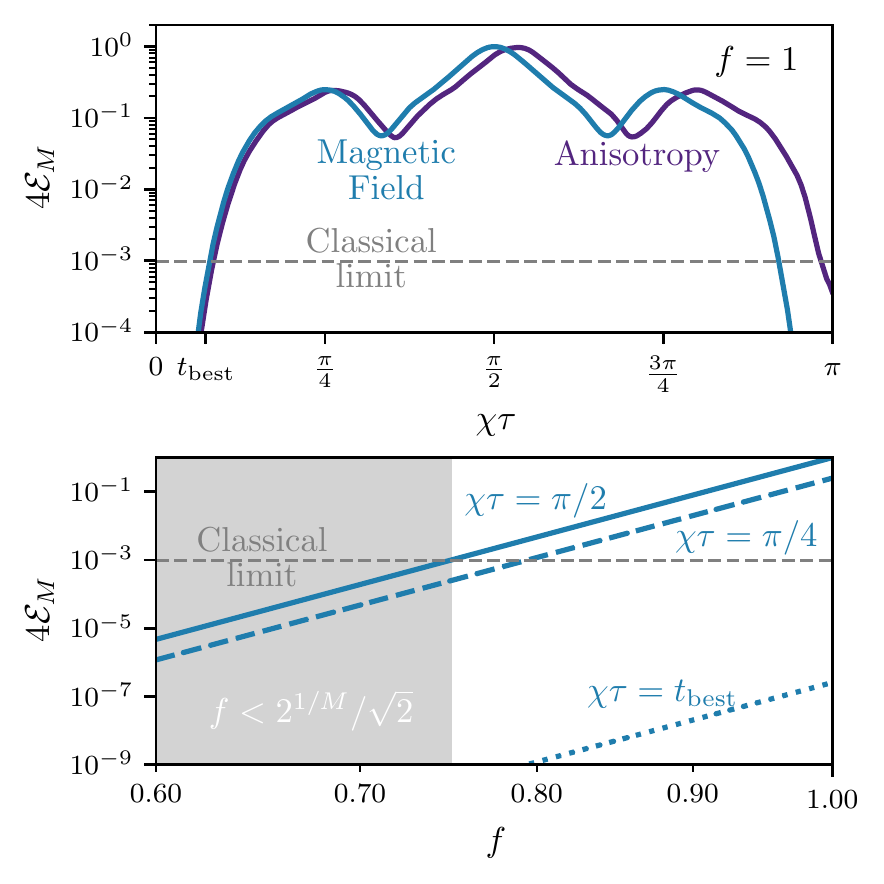}
    \caption{
   $M$-sites Bell correlator for the Hamiltonian (\ref{eq:t-J}) with $M=12$ sites under scenario A (Anisotropy) with $\Delta = 0.98, \beta_j = 0$ (blue line) and scenario B (Magnetic Field) with $\Delta = 1, \beta_j = J / 50 \cos\left[\pi (M-1)/M(j-1/2)\right]$ (orange lines). The upper panel shows the evolution of the correlator for $f=1$. The lower panel shows the scaling at three particular moments in time with the filling factor $f$ for scenario B. The solid line corresponds to the GHZ state, the dashed line to a superposition of $q=4$ coherent states, and the dotted line to the best spin-squeezing time $t_{\rm best}$.
   The logarithmic scale has been used to illustrate the power-law dependence of $f$ and the classical limit $2^{-M}$.
   }
    \label{fig:fig4}
\end{figure}

In the upper panel of Fig.~\ref{fig:fig4} we show the evolution of the $M$-site Bell correlator $\mathcal{E}_M$ with $M=12$ and for the two proposed methods, A and B. For the $M$-site correlator the only relevant contribution in (\ref{eq:coherentstate-holes}) comes from the part of the system state describing all sites occupied, as already discussed in Section \ref{sec: ToyModel}. The tunnelling of holes cannot change the value of this correlator.

In the lower panel of Fig.~\ref{fig:fig4} we check the scaling of the correlator with the filling factor $f$ for relevant instants in time, $\chi \tau=\pi/2$ (solid line) corresponding to the GHZ state. 
We see good agreement with the scaling of the toy model result in Eq.~(\ref{eq: corrTP0}) when identifying $p$ with $f$.  
Likewise, we see the GHZ state result crossing the classical limit at the value obtained in Eq.~(\ref{eq: p_c}) for $q=2$, identified by the edge of the grey area.

\subsection{Two-sites Bell correlations}
\label{subsec:2b-holes}

In the short-time dynamics, where spin squeezing is generated, the two-site Bell correlator $L$ given by (\ref{eq:FirstBell}) is more relevant.

\begin{figure}[]
    \centering
    \includegraphics{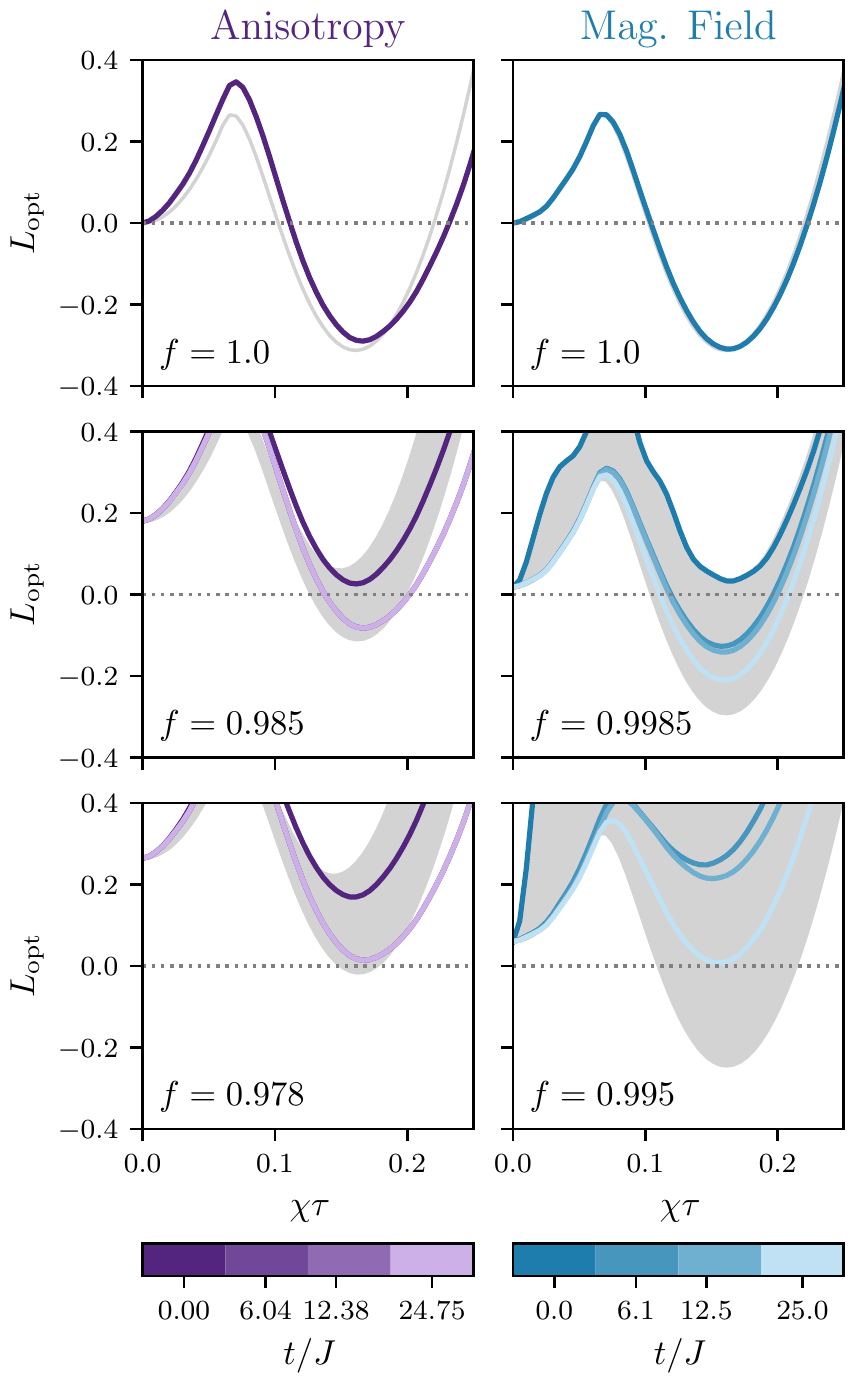}
    \caption{
    Evolution of the 
    two-sites Bell correlator (\ref{eq:FirstBell}) optimized over $\theta$ for the Hamiltonian (\ref{eq:t-J}) with $M = 12$ sites under the scenario A (Anisotropy) with $\Delta = 0.98, \beta_j = 0$ (left column) and scenario B (Mag. Field) with $\Delta = 1, \beta_j = J / 50 \cos\left[\pi (M-1)/M(j-1/2)\right]$ (right column).
    Each panel corresponds to a given filling factor $f$, as indicated in their lower left corners. Lines of different colours show different values of $t/J$ (see colour bars). 
    Grey areas indicate the regions between the lower and upper bounds as explained in the text, where $\chi$ is calculated with (\ref{eq:chi0}) or (\ref{eq:chi1}) assuming a single chain of size $M$. 
    }
    \label{fig:fig5}
\end{figure}

\begin{figure}
    \centering
    \includegraphics{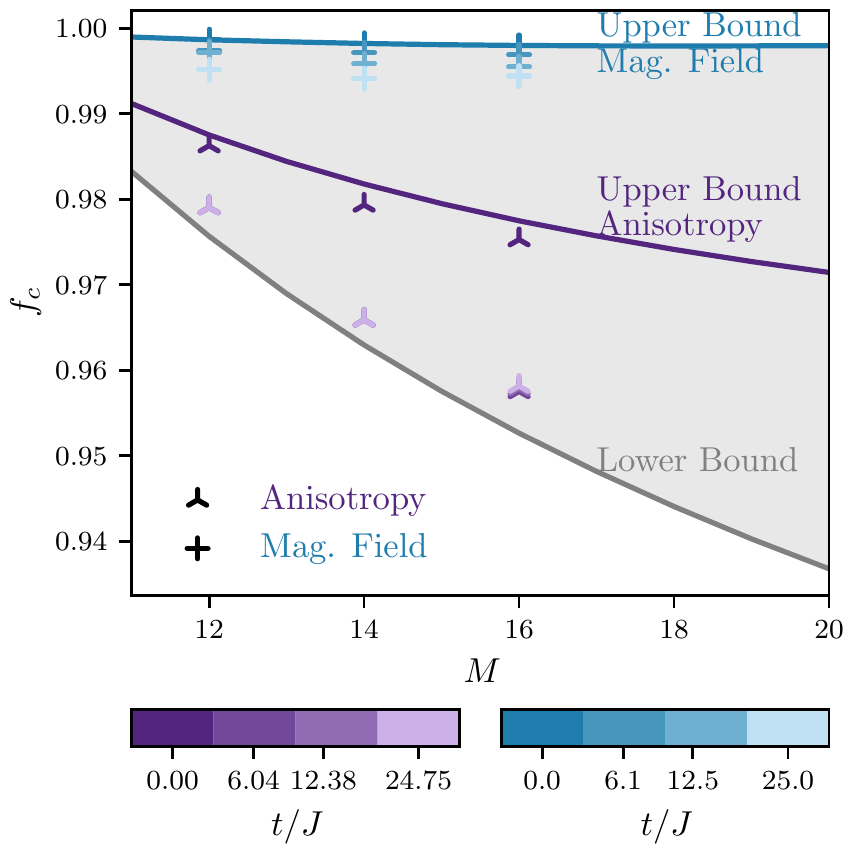}
    \caption{
    The critical value of filling factor $f_c$ concerning the system size $M$ for the Hamiltonian (\ref{eq:t-J}) under scenario A (Anisotropy) with $\Delta = 0.98, \beta_j = 0$ (triangular markers) and scenario B (Mag. Field) with $\Delta = 1, \beta_j = J / 50 \cos\left[\pi (M-1)/M(j-1/2)\right]$ (cross markers).
    The lower and upper bounds are shown in solid lines, with the area between them filled in grey. 
    We numerically find $f_c$ from similar data to the ones shown in Fig. \ref{fig:fig5}.
    Due to numerical calculation costs, results for $M=12$ include configurations of all possible numbers of holes in the initial state, while results for $M=14,16$ include configurations with at least three holes in (\ref{eq:longrhoa}).
    }
    \label{fig:bound_comparison}
\end{figure}

In Fig.~\ref{fig:fig5} we demonstrate the variation of the two-sites Bell correlations generated using the method A (interactions anisotropy, left column) and B (inhomogeneous magnetic field, right column) for different filling factors and effective tunnelling rates $t/J$.
The color bar indicates the magnitude of effective tunnelling when holes are present in the system relative to the relevant energy scale~$J$. 
While in all instances $t \ne 0$ to have a non-zero value of $J$, we assume for the effective tunnelling $t/J=0$ that positions of spins are fixed for those particular numerical simulations. 
Shaded areas indicate the upper and lower boundaries set by the results of holes fixed in place and holes moving infinitely fast, respectively. 
The upper bounds are calculated numerically for the case of the fixed position of holes by using effective models as described in \cite{PhysRevB.109.214310}. Note that the microscopic effective models (\ref{eq:chi0}) and (\ref{eq:eff_inhomo}) differ by the presence of a linear term, such that their results will be bounded by different limits as illustrated in Fig. \ref{fig:bound_comparison}. 
The lower bound is given by the toy model and the corresponding Bell inequality (\ref{eq:L_opt_pM}) where expectation values $\langle {\bf \cdot} \rangle_{\rm SS}$ are replaced with the one given by the corresponding OAT model.
We observe that the generation of Bell correlations is possible with a lower filling factor in the anisotropy method (A), 
both when the movement of holes is allowed and when not.
Only in this case, the numerical results 
demonstrate the effectiveness of the proposed toy model  (\ref{eq:L_opt_pM}) in the estimation of the lower bound in the detection of two-sites Bell correlations. 
In the case of inhomogeneous magnetic field method (B) we could not reach the lower bound within the explored region of parameters.

The critical value for the lower bound in this figure is obtained by solving the cubic equation obtained by setting (\ref{eq:L_opt_pM}) to zero, for which only one real root exists.
The anisotropy scenario A, while less accurately described by the corresponding effective OAT model (\ref{eq:chi0}) as explained in~\cite{PhysRevB.109.214310}, achieves the lower bound results predicted by the toy model~(\ref{eq:L_opt_pM}) for a sufficiently large ratio $t/J=U_{ab}/4t$.
This is much harder to achieve in the inhomogeneous magnetic field scenario B since the linear term in (\ref{eq:eff_inhomo}) changes at each configuration of holes and makes the dynamics largely incoherent.
This is a stark contrast with the M-site Bell correlator presented in Fig.~\ref{fig:fig4} where we find the lower bound set by the toy model~(\ref{eq: corrTP0}) to be accurate for the two scenarios.

\section{Entanglement generation in the superfluid phase}\label{sec: SF}

We consider here the protocol where the quantum correlations are generated using atom-atom interactions within the superfluid phase and then stored in the Mott phase through adiabatic increasing of the lattice height \cite{KajtochEPL2018,PlodzienPRA2020}. 
We assume that the tunnelling parameter $t$ and interaction parameters $U_{\sigma\sigma'}$ can by approximated as \cite{PlodzienPRA2020}
\begin{align}
    t &\approx \frac{4}{\sqrt{\pi}} \left(\frac{V_0}{E_r}\right)^{3/4} e^{-2\sqrt{{V_0}/{E_r}}},\\
    U_{\sigma\sigma'} &\approx \sqrt{\frac{32}{\pi}} \frac{a_{\sigma\sigma'}d}{L_{\perp}^2}\left(\frac{V_0}{E_r}\right)^{1/4},
\end{align}
where $a_{\sigma\sigma'}$ are the s-wave scattering lengths, $V_0$ is the lattice potential depth and $L_{\perp}$ is the characteristic length of the wave function perpendicular to the lattice direction.

In this protocol, initially, all atoms occupy the internal state $a$, and at zero temperature, they are in the ground state of the Bose-Hubbard Hamiltonian $|\psi^{(a)}_0\rangle$ in the superfluid regime. 
Subsequently, a $\pi/2$ pulse is applied to put the atoms in a coherent spin state where the collective spin is along the $x$-axis. During the dynamics, the lattice height $V_0$ linearly increases as
\begin{align}\label{eq: linramp}
V_0(\tau)=V_{\rm ini}+(V_{\rm fin}-V_{\rm ini})\frac{\tau}{\tau_\mathrm{ramp}},
\end{align}
such that, at $\tau=\tau_\mathrm{ramp}$, for an adiabatic evolution and initially zero temperature, the system reaches the Mott-squeezed state where the atoms are spatially distributed with one atom per lattice site and show entanglement in their internal degree of freedom (from spin squeezed state up to a GHZ state according to the chosen ramp duration $\tau_\mathrm{ramp}$). This corresponds to a state of the form (\ref{eq:system-state}) with $\hat\rho_{\rm ext}=\bigotimes_{i=1}^M|1\rangle_j {}_j\langle 1|$.  At finite temperature $T$, the system exhibits initial thermal fluctuations described by the density operator
\begin{align}\label{eq: MixtState}
\hat\rho_T=\frac{1}{Z}\sum_n e^{-E_n^{(0)}/k_{\rm B}T}|\psi_n^{(a)}\rangle\langle\psi_n^{(a)}|,
\end{align} 
where $k_{\rm B}$ is the Boltzmann constant, $E_n^{(0)}$ are the eigenenergies of the initial Hamiltonian (in the superfluid phase), all atoms are in the internal mode $a$ so $|\psi_n^{(a)}\rangle$ are the corresponding eigenstates, and $Z=\sum_ne^{-E_n^{(0)}/k_{\rm B}T}$ is the normalization constant. These initial thermal fluctuations lead, at the end of the ramp, to a non-zero probability of having holes and double occupations of the lattice sites. In this section, we explore the effect of these initial thermal fluctuations on the detection of the $M$-site and two-sites non-local Bell correlations present in the final state. We first note that numerical observations performed in small $1D$ lattices reveal that when $U_{aa}=U_{bb}\gtrsim U_{ab}> 0$, which is relevant to our purposes, the external dynamics is weakly affected by the internal dynamics up to the Mott transition. Consequently, we estimate the occupation statistics of different lattice sites in the final state by restricting the Bose-Hubbard model to a single internal state $a$. In this case, the spectrum of the Hamiltonian in the deep Mott phase with $t\to 0$, is simple: a non-degenerate ground state $|\psi_{0}^{\rm MI}\rangle$, with one atom per lattice site, and gapped excited states $|\psi_{n}^{\rm MI}\rangle$ showing holes and doubly occupied sites. The ground state $|\psi_{0}^{\rm MI}\rangle$ is obtained, from the initial ground superfluid state $|\psi_{0}^{(a)}\rangle$, by a unitary evolution with the time dependent Bose-Hubbard Hamiltonian
\begin{align}
|\psi_{0}^{\rm MI}\rangle=\hat U|\psi_{0}^{(a)}\rangle.
\end{align}

\subsection{$M$-site Bell correlations}

\begin{figure}[]
    \centering
    \includegraphics[width=0.48\textwidth]{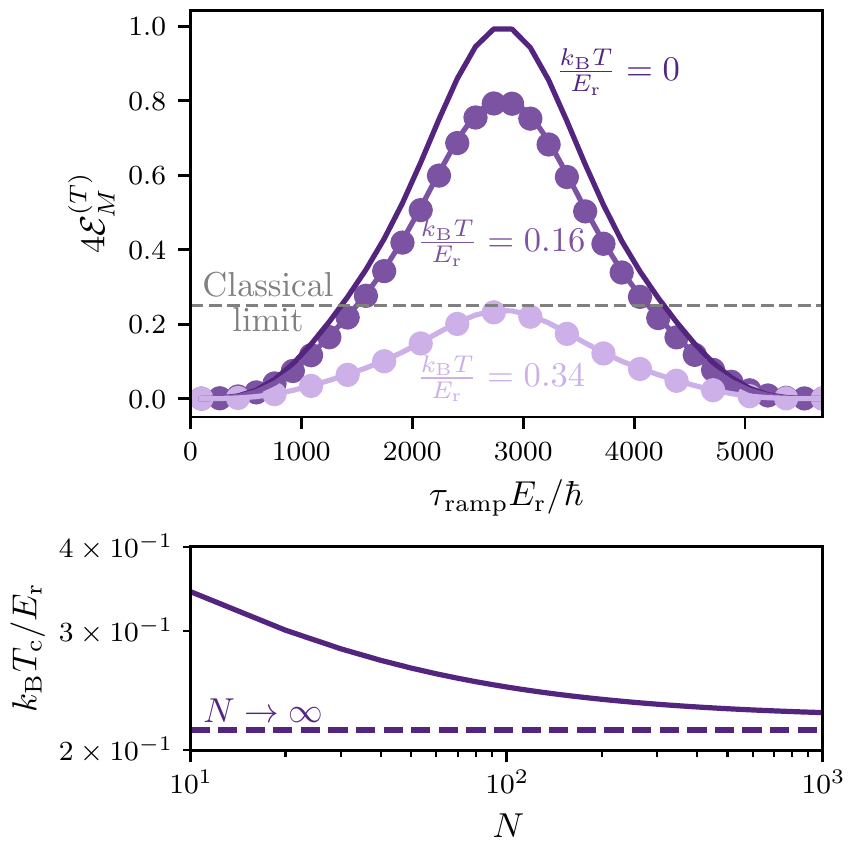}
    \caption{Upper panel: $M$-sites Bell correlator $\mathcal{E}_M^{(T)}$ at zero and nonzero temperature as a function of the ramp duration $\tau_\mathrm{ramp}$ for $N=M=4$ with initial lattice height $V_{\rm ini}/E_{\rm r}=3$ and final lattice height $V_{\rm fin}/E_{\rm r}=20$. Parameters used here are: $a_{a}=a_{b}=100.4 a_{\rm B}$ and $a_{ab}=0.95 a_{a}$, with $a_{\rm B}$ being the Bohr radius, $d = 431$ nm, $L_{\perp}=d / \sqrt{2\pi}$. Colored points represent the exact numerical results, including the $\pi/2$ pulse and the dynamical raising of the lattice, and the colored solid lines superimposed to the symbols represent Equation (\ref{eq: CorrT}) where $P_0$ was calculated using exact numerical simulations. The horizontal grey dashed line represents the non-locality bound. Lower panel: Critical temperature $T_{\rm c}$, below which the non-local Bell correlations associated with the GHZ state can be detected, as a function of atom number $N$. The dashed line corresponds to the large $N$ limit (\ref{eq:large_N_T_M_body}).
    }
    \label{fig: corrbell}
\end{figure}

Since the excited states $|\psi_n^{\rm MI}\rangle$ have at least one hole and one doubly occupied lattice site, they give no contribution to the $M$-site Bell correlator (\ref{eq: corrTP0}) whose value at non-zero temperature is given by
\begin{align}\label{eq: CorrT}
\mathcal{E}_M^{(T\ne0)}=P_0^2\mathcal{E}_M^{(T=0)},
\end{align}
where $P_0=e^{-E_0^{(0)}/k_{\rm B}T}/Z$ represents the probability of occupying the ground state in (\ref{eq: MixtState})\footnote{It is important to note that in general, the final external configuration of the system does not consist of a factorized state over lattice sites, as in (\ref{eq:extern-state}), in particular the probability $P_0$ is not of the form $P_0=p^M$.}. 
The upper panel of Fig. \ref{fig: corrbell} illustrates this relationship for $N=M=4$ over varying temperatures, revealing the existence of a critical temperature $T_{\rm c}$ below which $M$-site Bell correlations are detectable. In the weakly interacting regime, the probability $P_0$ can be analytically determined using the Bogoliuobov theory, where, in the initial superfluid regime, when all the atoms occupy the internal state $a$, the system can be approximately described by the Hamiltonian
\begin{align}\label{eq: Bog_Model}
\hat H_{\rm Bog}=E_0+\sum_{\substack{j\neq 0}}\hbar\omega_j\hat d_j^{\dag}\hat d_j.
\end{align}
In equation (\ref{eq: Bog_Model}), $E_0$ is the ground state energy of the system, $\hat d_j$ is the annihilation operator of a Bogoliubov quasi-particle associated with the quasi-momentum $q_j$, and $\hbar\omega_j$ is given in terms of the tunneling parameter $t$ and the interaction parameter $U=U_{aa}$, by
\begin{align}\label{eq: Bog_spectr}
\hbar\omega_j=4t\sqrt{\sin^2(\frac{\pi}{N}j)\left(\sin^2(\frac{\pi}{N}j)+\frac{U}{2t}\frac{N}{M}\right)}.
\end{align}
At non-zero temperature $T$, the probability $P_0$ of being in the ground state is given by 
\begin{align}\label{eq: ProbGS}
P_0=\prod_{j\neq0}p_j(n_j=0),
\end{align}
where, $p_j(n_j=n)=e^{-n\hbar\omega_j/k_{\rm B}T}/Z_j$ with $Z_j=(1-e^{-\hbar\omega_j/k_{\rm B}T})^{-1}$. By replacing in (\ref{eq: ProbGS}), one obtains
\begin{align}\label{eq: P0Bog}
P_0=\prod_{\substack{j\neq 0}}\left(1-e^{-\hbar\omega_j/k_{\rm B}T}\right).
\end{align}
The critical temperature $T_{\rm c}$, below which Bell correlations in states generated at times $\chi \tau=\pi/q$ of OAT dynamics are detectable, can be determined by setting $P_0$ (\ref{eq: P0Bog}) equal to the critical probability $P_0=p_{\rm c}^M$, where $p_{\rm c}$ is given by (\ref{eq: p_c}). The lower panel of Fig. \ref{fig: corrbell} illustrates this critical temperature for a GHZ state (i.e. $q=2$) as a function of the atom number $N$.
One can find analytically the critical temperature for $N\to\infty$ by taking the continuous limit in (\ref{eq: P0Bog}), which yields
\begin{equation}
\begin{split}
    \ln{P_0} &\simeq \int_{-\infty}^{\infty} dj \ln\left({1-e^{-\hbar\omega_j/k_{\rm B}T}}\right)\\ 
    &\simeq - \frac{\pi k_\mathrm{B}T N}{6\sqrt{2tU}}.
\end{split}
\end{equation}
We again take $P_0 = p_c^M$ for $q=2$ to obtain the critical temperature 
\begin{equation}
    \label{eq:large_N_T_M_body}
    k_\mathrm{B} T_c = \frac{3\ln{2}}{\pi}\sqrt{2 t U},
\end{equation}
which is illustrated as a dashed line in the lower panel of Fig. \ref{fig: corrbell}.

\subsection{Two-sites Bell correlations}

Here, we focus on two-site correlations that can be measured also in large systems.
In this context, we assume that the system is brought to the Mott phase through a transformation that leaves the system in thermal equilibrium at each moment and conserves the entropy. Under such conditions, one can explicitly write an approximate external density matrix $\hat\rho_{\rm ext}$ of the system in the final Mott phase and thus analytically determine the probability $p(T_{\rm i})$ of having exactly one atom \textit{in a given lattice site} as a function of the initial temperature $T_{\rm i}$. We first calculate the initial entropy $S_{\rm SF}$ as a function of $T_{\rm i}$. We then determine the final temperature $T_{\rm f}$ of the system in Mott using entropy conservation \cite{BlakiePRA2004,AnaMRPRA2006,YoshimuraPRA2008}
\begin{align}\label{eq: Entropy_conservation}
S_{\rm SF}(T_{\rm i})|_{V_0=V_{\rm ini}}=S_{\rm Mott}(T_{\rm f})|_{V_0=V_{\rm fin}}.
\end{align} 
In the weakly interacting regime, in the superfluid phase, the entropy can be calculated using the Bogoliubov theory. Indeed, By using the Bogoliubov spectrum (\ref{eq: Bog_spectr}), the system partition function at temperature $T$ can be written as (non interacting bosons)
\begin{align}
Z_{\rm Bog}(T)=\prod_j\left(1-e^{-\frac{\hbar\omega_j}{k_{\rm B}T}}\right)^{-1}.
\end{align}
The free energy defined as $F_{\rm Bog}=-k_{\rm B}T\ln Z_{\rm Bog}$ is given by
\begin{align}
F_{\rm Bog}(T)=k_{\rm B}T\sum_j\ln(1-e^{-\frac{\hbar\omega_j}{k_{\rm B}T}}).
\end{align}
We thus deduce the entropy of the system as
\begin{align}
&\frac{S_{\rm SF}^{(\rm Bog)}(T)}{k_{\rm B}} \equiv-\partial_T F/k_{\rm B}\nonumber \\
&=-\sum_j\left[\ln(1-e^{-\frac{\hbar\omega_j}{k_{\rm B}T}})+\frac{\hbar\omega_j}{k_{\rm B}T}\frac{e^{-\frac{\hbar\omega_j}{k_{\rm B}T}}}{1-e^{-\frac{\hbar\omega_j}{k_{\rm B}T}}}\right].
\end{align}

We now calculate the entropy of the system in the Mott phase by considering the limit $t \to 0$ and small temperatures. In this case, the partition function, when $N=M$, can be estimated using a two particles-holes excitation approximation, where, we take into account only the states with at most two holes and two doubly occupied lattice sites, with $U=U_{aa}$
\begin{align}\nonumber
Z_{\rm Mott}(T)&=1+M(M-1)e^{-\frac{U}{k_{\rm B}T}} \\
&\quad+M(M-1)(M-2)(M-3)e^{-\frac{2U}{k_{\rm B}T}}.
\end{align} 
By using the free energy, in the Mott phase, one can obtain the system entropy
\begin{align}\nonumber
\frac{S_{\rm Mott}(T)}{k_{\rm B}}&=\ln Z_{\rm Mott}+\frac{1}{Z_{\rm Mott}}\frac{U}{k_{\rm B}T}\left[M(M-1)e^{-\frac{U}{k_{\rm B}T}}\right.\\
&\quad\left.+M(M-1)(M-2)(M-3)e^{-\frac{2U}{k_{\rm B}T}}\right].
\end{align} 
After determining $T_{\rm f}$ using (\ref{eq: Entropy_conservation}), we approximate the external density matrix of the system at the end of the ramp as 
\begin{align}\nonumber
\hat\rho_{\rm ext}&
\approx
\frac{1}{Z_{\rm Mott}(T_{\rm f})}
\left[|\psi_0^{0\rm h}\rangle\langle\psi_0^{0\rm h}|
\right. \\
{}&\left.
+e^{-\frac{U}{k_{\rm B}T_{\rm f}}}\sum_{j=1}^M\sum_{k\neq j}^M|\psi_{jk}^{1\rm h}\rangle\langle\psi_{jk}^{1\rm h}|\right.\nonumber  \\
&\quad\left.+e^{-\frac{2U}{k_{\rm B}T_{\rm f}}}\sum_{j=1}^M\sum_{k\neq j}^M\sum_{l\neq j,k}^M\sum_{m\neq j,k,l}^M|\psi_{jklm}^{2\rm h}\rangle\langle\psi_{jklm}^{2\rm h}|\right],
\end{align} 
where $|\psi_0^{0\rm h}\rangle$ is the state with exactly one atom per lattice site (zero holes), $|\psi_{jk}^{1\rm h}\rangle$ is the state with only one hole in the $j^{\rm th}$ site and a single double occupancy in the $k^{\rm th}$ site and $|\psi_{jklm}^{2\rm h}\rangle$ is the state with two holes in the $j^{\rm th}$ and $k^{\rm th}$ sites respectively and two double occupancies in the $l^{\rm th}$ and $m^{\rm th}$ sites respectively. This approximation enables us to analytically calculate, as a function of the initial temperature, the probability $p$ of having a Fock state with one atom in a given lattice site
\begin{align}\label{eq: pr_Fock_1}\nonumber
p&\equiv{\rm tr}\left[|1\rangle_i{}_i\langle1|\,\hat\rho_{\rm ext}\right]\\\nonumber
&=\frac{1}{Z_{\rm Mott}(T_{\rm f})}\left[1+(M-1)(M-2)e^{-\frac{U}{k_{\rm B}T_{\rm f}}}\right.\\
&\quad\left.+(M-1)(M-2)(M-3)(M-4)e^{-\frac{2U}{k_{\rm B}T_{\rm f}}}\right]
\end{align} 

\begin{figure}[]
	\centering
	\includegraphics[width=0.49\textwidth]{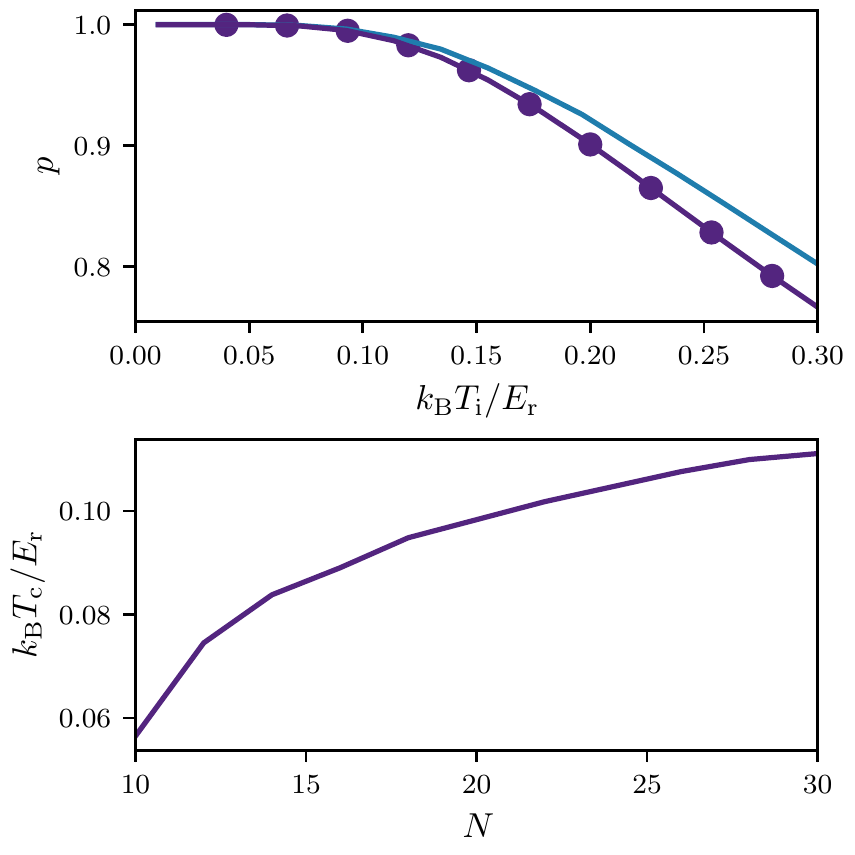}
	\caption{Upper panel: Probability $p$ of having a Fock state with one atom in a given site at the end of the ramp for $N=M=4$ as a function of the initial temperature $T_{\rm i}$ given by (\ref{eq: pr_Fock_1}), in solid blue, where the final temperature $T_{\rm f}$ was calculated using (\ref{eq: Entropy_conservation}) where the superfluid entropy $S_{\rm SF}(T_{\rm i})$ was calculated using exact numerical simulations. This is compared to the exact numerical results, with $V_{\rm ini}/E_{\rm r}=3$ and $V_{\rm fin}/E_{\rm r}=30$, calculated using, in purple points, the one-component Bose-Hubbard Hamiltonian and, in solid purple, the two-components Bose-Hubbard Hamiltonian. 
    Lower panel: Critical temperature $T_{\rm c}$, associated to the minimal value of $p_{\rm c}$, above which two-sites non-local Bell correlations cannot be detected, as a function of the number of atoms $N$.}
	\label{fig:Critic_T}
\end{figure} 

In numerical simulations of the exact dynamics, $p$ can be calculated as the trace of the projection of the density matrix of the system on the eigenspace of the observable $\hat n_i$ associated with the eigenvalue $n_i=1$. Equation (\ref{eq: pr_Fock_1}) is represented in the upper panel of Fig. \ref{fig:Critic_T} as a function of the initial temperature, compared to exact results using both one- and two-component Bose-Hubbard Hamiltonian. This reveals that at sufficiently low temperatures in small systems, the probability $p(T_{\rm i})$ can be accurately approximated using (\ref{eq: pr_Fock_1}) and the entropy conservation condition (\ref{eq: Entropy_conservation}). By equating $p(T_{\rm i})$ with the critical probability (\ref{eq:pcr2data}) from the toy model, one can determine an upper bound, dependent on the ramp duration $\tau_\mathrm{ramp}$, for the initial temperature $T_{\rm c}$ above which two-sites non-local Bell correlations cannot be detected. The lower panel of Fig. \ref{fig:Critic_T} shows the critical temperature associated to the smallest critical $p$, minimized over the ramp duration $\tau_\mathrm{ramp}$, as a function of the atom number $N$~
\footnote{For larger atoms number, beyond the range of $N$ presented in the figure, the two-hole approximation becomes invalid.}.

\section{Summary and conclusions}
\label{sec:summ}

In this paper, we considered the detection of Bell correlations using two-level ultra-cold bosonic atoms loaded into optical lattices. 
We focus on identifying Bell violations for dynamically generated entangled states when imperfections related to non-unit filling appear.

Our proposed toy model accounts for imperfections that provide non-unit filling per site in the preparation and measurement stages through the probability $p$ of a site being single-occupied. The $M$-site correlator shows a violation of the Bell inequality for GHZ states when $p>p_c=2^{1/M}/\sqrt{2}$. 
On the other hand, the two-site Bell correlator relying on collective spin measurements for spin-squeezed states shows the bounds $p_c=4/5$ when $N=pM<M$, and $p_c=\sqrt{3}/2$ when $N=M$. 
We illustrate these general results using two different methods of entanglement generation implying different sources of imperfections: vacant sites due to initial non-unit filling, and vacant and multiply occupied sites due to non-zero temperature of the initial state. 
In the presence of holes, we study the generation of entanglement in the Mott insulating via interaction anisotropy and inhomogeneous field demonstrating the validity of the toy model predictions for the critical value of the filling factor allowing the detection of violation of Bell inequalities.
For non-zero temperatures, we study the entanglement generation in the superfluid regime produced via atom-atom interactions transferred to the Mott regime via adiabatic rising of the lattice height. 
We find a connection between the probability $p$, of a single occupation of a given site, of the toy model and the effective temperature $T$ of the initial state. 
We then identified the critical value of the initial temperature allowing violation of Bell inequalities.

Our results reveal the fundamental limits on detecting Bell correlations in the lattice system due to occupation defects. 
Analytical predictions of the toy model are general and can be relevant for any platform where addressing individual spins is possible.
The bounds identified in specific protocols fall within the range of typical experimental realizations, suggesting the feasibility of detecting Bell correlations.

\section*{ACKNOWLEDGMENTS}

We gratefully acknowledge discussions with Ir\'{e}n\'{e}e Fr\'{e}rot. This work was supported by the Polish National Science Center DEC-2019/35/O/ST2/01873 and DEC-2023/48/Q/ST2/00087. 
T.H.Y acknowledges support from the Polish National Agency for Academic Exchange through the Foreign Doctoral Internship Grant NAWA Preludium BIS 1 No. PPN/STA/2021/1/00080. 
A part of the computations was done at the Centre of Informatics Tricity Academic Supercomputer \& Network.

\bibliography{biblio}

\end{document}